\documentclass[twocolumn,secnumarabic,amssymb, nobibnotes, aps, prl,
superscriptaddress, figure ]{revtex4-2}

\usepackage{natbib}
\usepackage[dvipdfmx,hiresbb]{graphicx}
\begin{document}

\title{Rabi-Oscillation Spectroscopy of the Hyperfine Structure
of Muonium Atoms}%

\author{S.~Nishimura}%
\email[]{shoichiro.nishimura@kek.jp}
\affiliation{High Energy Accelerator Research Organization (KEK), 1-1 Oho, Tsukuba, Ibaraki 305-0801, Japan}
\affiliation{J-PARC Center, 2-4 Shirakata, Tokai, Ibaraki 319-1195, Japan}

\author{H.~A.~Torii}%
\affiliation{School of Science, The University of Tokyo, 7-3-1 Hongo, Bunkyo, Tokyo 113-0033, Japan}


\author{Y.~Fukao}%
\affiliation{High Energy Accelerator Research Organization (KEK), 1-1 Oho, Tsukuba, Ibaraki 305-0801, Japan}
\affiliation{J-PARC Center, 2-4 Shirakata, Tokai, Ibaraki 319-1195, Japan}
\affiliation{The Graduate University for Advanced Studies (SOKENDAI), 1-1 Oho, Tsukuba, Ibaraki 305-0801, Japan}



\author{T.~U.~Ito}%
\affiliation{J-PARC Center, 2-4 Shirakata, Tokai, Ibaraki 319-1195, Japan}
\affiliation{Advanced Science Research Center, Japan Atomic Energy Agency, 2-4 Shirakata, Tokai, Ibaraki 319-1195, Japan}

\author{M.~Iwasaki}%
\affiliation{RIKEN, 2-1 Hirosawa, Wako, Saitama 351-0198, Japan}



\author{S.~Kanda}%
\affiliation{RIKEN, 2-1 Hirosawa, Wako, Saitama 351-0198, Japan}

\author{K.~Kawagoe}%
\affiliation{Kyushu University, 744 Motooka, Nishi, Fukuoka 819-0395, Japan}

\author{D.~Kawall}%
\affiliation{University of Massachusetts Amherst, 1126 Lederle Graduate Research Tower, Amherst, MA 01003-9337, USA}

\author{N.~Kawamura}%
\affiliation{High Energy Accelerator Research Organization (KEK), 1-1 Oho, Tsukuba, Ibaraki 305-0801, Japan}
\affiliation{J-PARC Center, 2-4 Shirakata, Tokai, Ibaraki 319-1195, Japan}
\affiliation{The Graduate University for Advanced Studies (SOKENDAI), 1-1 Oho, Tsukuba, Ibaraki 305-0801, Japan}




\author{N.~Kurosawa}%
\affiliation{High Energy Accelerator Research Organization (KEK), 1-1 Oho, Tsukuba, Ibaraki 305-0801, Japan}
\affiliation{J-PARC Center, 2-4 Shirakata, Tokai, Ibaraki 319-1195, Japan}

\author{Y.~Matsuda}%
\affiliation{Graduate School of Arts and Sciences, The University of Tokyo, 3-8-1 Komaba, Meguro, Tokyo 153-8902, Japan}

\author{T.~Mibe}%
\affiliation{High Energy Accelerator Research Organization (KEK), 1-1 Oho, Tsukuba, Ibaraki 305-0801, Japan}
\affiliation{J-PARC Center, 2-4 Shirakata, Tokai, Ibaraki 319-1195, Japan}
\affiliation{The Graduate University for Advanced Studies (SOKENDAI), 1-1 Oho, Tsukuba, Ibaraki 305-0801, Japan}

\author{Y.~Miyake}%
\affiliation{High Energy Accelerator Research Organization (KEK), 1-1 Oho, Tsukuba, Ibaraki 305-0801, Japan}
\affiliation{J-PARC Center, 2-4 Shirakata, Tokai, Ibaraki 319-1195, Japan}
\affiliation{The Graduate University for Advanced Studies (SOKENDAI), 1-1 Oho, Tsukuba, Ibaraki 305-0801, Japan}



\author{N.~Saito}%
\affiliation{High Energy Accelerator Research Organization (KEK), 1-1 Oho, Tsukuba, Ibaraki 305-0801, Japan}
\affiliation{J-PARC Center, 2-4 Shirakata, Tokai, Ibaraki 319-1195, Japan}
\affiliation{The Graduate University for Advanced Studies (SOKENDAI), 1-1 Oho, Tsukuba, Ibaraki 305-0801, Japan}
\affiliation{School of Science, The University of Tokyo, 7-3-1 Hongo, Bunkyo, Tokyo 113-0033, Japan}

\author{K.~Sasaki}%
\affiliation{High Energy Accelerator Research Organization (KEK), 1-1 Oho, Tsukuba, Ibaraki 305-0801, Japan}
\affiliation{J-PARC Center, 2-4 Shirakata, Tokai, Ibaraki 319-1195, Japan}
\affiliation{The Graduate University for Advanced Studies (SOKENDAI), 1-1 Oho, Tsukuba, Ibaraki 305-0801, Japan}

\author{Y.~Sato}%
\affiliation{High Energy Accelerator Research Organization (KEK), 1-1 Oho, Tsukuba, Ibaraki 305-0801, Japan}

\author{S.~Seo}%
\affiliation{RIKEN, 2-1 Hirosawa, Wako, Saitama 351-0198, Japan}
\affiliation{Graduate School of Arts and Sciences, The University of Tokyo, 3-8-1 Komaba, Meguro, Tokyo 153-8902, Japan}

\author{P.~Strasser}%
\affiliation{High Energy Accelerator Research Organization (KEK), 1-1 Oho, Tsukuba, Ibaraki 305-0801, Japan}
\affiliation{J-PARC Center, 2-4 Shirakata, Tokai, Ibaraki 319-1195, Japan}
\affiliation{The Graduate University for Advanced Studies (SOKENDAI), 1-1 Oho, Tsukuba, Ibaraki 305-0801, Japan}

\author{T.~Suehara}%
\affiliation{Kyushu University, 744 Motooka, Nishi, Fukuoka 819-0395, Japan}

\author{K.~S.~Tanaka}%
\affiliation{Tohoku University, 2-1-1 Katahira, Aoba, Sendai, Miyagi 980-8577, Japan}

\author{T.~Tanaka}%
\affiliation{RIKEN, 2-1 Hirosawa, Wako, Saitama 351-0198, Japan}
\affiliation{Graduate School of Arts and Sciences, The University of Tokyo, 3-8-1 Komaba, Meguro, Tokyo 153-8902, Japan}

\author{J.~Tojo}%
\affiliation{Kyushu University, 744 Motooka, Nishi, Fukuoka 819-0395, Japan}



\author{A.~Toyoda}%
\affiliation{High Energy Accelerator Research Organization (KEK), 1-1 Oho, Tsukuba, Ibaraki 305-0801, Japan}
\affiliation{J-PARC Center, 2-4 Shirakata, Tokai, Ibaraki 319-1195, Japan}
\affiliation{The Graduate University for Advanced Studies (SOKENDAI), 1-1 Oho, Tsukuba, Ibaraki 305-0801, Japan}


\author{Y.~Ueno}%
\affiliation{RIKEN, 2-1 Hirosawa, Wako, Saitama 351-0198, Japan}

\author{T.~Yamanaka}%
\affiliation{Kyushu University, 744 Motooka, Nishi, Fukuoka 819-0395, Japan}

\author{T.~Yamazaki}%
\affiliation{High Energy Accelerator Research Organization (KEK), 1-1 Oho, Tsukuba, Ibaraki 305-0801, Japan}
\affiliation{J-PARC Center, 2-4 Shirakata, Tokai, Ibaraki 319-1195, Japan}
\affiliation{The Graduate University for Advanced Studies (SOKENDAI), 1-1 Oho, Tsukuba, Ibaraki 305-0801, Japan}

\author{H.~Yasuda}%
\affiliation{School of Science, The University of Tokyo, 7-3-1 Hongo, Bunkyo, Tokyo 113-0033, Japan}

\author{T.~Yoshioka}%
\affiliation{Kyushu University, 744 Motooka, Nishi, Fukuoka 819-0395, Japan}

\author{K.~Shimomura}%
\affiliation{High Energy Accelerator Research Organization (KEK), 1-1 Oho, Tsukuba, Ibaraki 305-0801, Japan}
\affiliation{J-PARC Center, 2-4 Shirakata, Tokai, Ibaraki 319-1195, Japan}
\affiliation{The Graduate University for Advanced Studies (SOKENDAI), 1-1 Oho, Tsukuba, Ibaraki 305-0801, Japan}

\collaboration{MuSEUM collaboration}

\date{February 2021}%

 \begin{abstract}
As a new method to determine the resonance frequency, Rabi-oscillation spectroscopy has been developed. 
In contrast to the conventional spectroscopy which draws the resonance curve, 
Rabi-oscillation spectroscopy fits the time evolution of the Rabi oscillation. 
By selecting the optimized frequency, it is shown that the precision is twice as good as the conventional spectroscopy with a frequency sweep. 
Furthermore, the data under different conditions can be treated in a unified manner,
 allowing more efficient measurements for systems consisting of a limited number of short-lived particles produced by accelerators such as muons.
We have developed a fitting function that takes into account the spatial distribution of muonium 
and the spatial distribution of the microwave intensity to apply the new method to ground-state muonium hyperfine structure measurements at zero field. 
This was applied to the actual measurement data and the resonance frequencies were determined under various conditions. 
The result of our analysis gives 
$\nu_{\rm HFS}=4\ 463\ 301.61 \pm 0.71\ {\rm kHz}$, which is the world's highest precision under zero field conditions.

 \end{abstract}
\maketitle

Atomic spectroscopy is a reliable tool for precision studies
of the properties of the elementary particles.
Measurement of the transition energy of an atom undergoing
electric and magnetic transitions can provide information on
the mass and magnetic moment of the particle constituting the atom.
Conventional spectroscopy plots signal intensity as a function of
a frequency swept over the range of the resonance frequency.
The resonance frequency is obtained by finding the center of the curve shaped by fitting an appropriate resonance function to the resonance curve.
In this paper, we introduce a new analysis method
where the resonance frequency can be obtained by fitting the simulated function to the
time evolution of the Rabi oscillation directly
without any frequency sweep. 
This method, named Rabi-oscillation spectroscopy, can essentially eliminate systematic uncertainties due to power fluctuations.
We report the first application of this method
to the spectroscopy of the spin resonance 
of muonium atoms.

Muonium is the bound state comprised of a positive muon and an electron, being as such one of several hydrogen-like atoms. 
Spectroscopy of muonium atoms is a promising method in the search for new physics
in particle physics research.
In the Standard Model of particle physics,
the positive muon and the electron are point-like lepton particles; therefore,
the contribution of the strong interaction is relatively small and well understood 
compared to hydrogen.
The muon-to-electron mass ratio can be found
from the muonium hyperfine structure (HFS)~\cite{RevModPhys.88.035009}.
The electron mass has been measured precisely, so the spectroscopy of
the muonium HFS provides the most precise estimation of the muon mass,
which is an essential input parameter for determinations of the Standard Model parameters,
such as the Fermi coupling constant.
Also, the precise measurement of the muonium HFS is important for the determination
of the muon anormalous magnetic moment, $g-2$~\cite{PhysRevD.101.014029, 10.1093ptepptz030, grange2015muon}.

In the previous experiments~\cite{PhysRevA.8.86,CASPERSON1975397,PhysRevLett.82.711},
the muonium HFS was measured by sweeping either the frequency of the microwave field or the strength of the external magnetic field.
The dominant source of the measurement uncertainty was the statistical nature of the experiment.
It is necessary to ensure the most efficient use of the data.
Considering that the microwave power can change during the frequency sweep, 
it will be an effective use of data if we no longer need to limit the data to those with stable conditions.

The complexity of the muonium resonance
experiment stems from the fact that a muon decays in a very short lifetime of about
$2.2\ \mu$s. 
In the conventional method, the muonium HFS interval is determined
by drawing a resonance curve of the time-integrated signal 
as a function of the microwave frequency; the width of the resonance curve limits the obtainable precision of the resonance frequency.
The key to reaching a higher precision is to reduce 
the resonance width, attributed to the natural width and power broadening.

In general, in the case of the homogeneous natural width or power broadening,
the resonance frequency can be obtained precisely by fitting a Lorentzian function to the resonance curve.
Even if an ideal resonance curve is constructed, 
it is still challenging to determine the resonance center frequency 
with an uncertainty smaller than one hundredth of the resonance width.
If there is inhomogeneous broadening, such as a Doppler width
or a particular systematic factor that makes 
the resonance curve asymmetric,
the precision of the resonance frequency measurement becomes even lower.
In particular, asymmetry in the microwave power across a resonance line would lead to difficulties 
in extracting the line center.
The natural width of the muonium resonance line is 145~kHz
due to the finite lifetime of the muon, so even achieving 1~kHz precision
on a line center is difficult.

One technique to improve the measurement precision is to reduce the resonance width.
It is possible to extend an apparent lifetime by using the time information of the signal.
Namely, the width in the vicinity of the resonance center becomes narrower 
by selecting the later signals, i.e., signals from those muonium atoms 
that have interacted for a longer period of time
with the microwave.
However, this method leaves the envelope curve wide because of
the side peaks.
Also, events occurring in later times are selected at the cost of the statistics.
Therefore, it is necessary to select an optimal time range
to reconcile the advantages of the narrower width of the center peak
with the disadvantages of limited statistics.
In an experiment where the time range is not selected by hardware, but
all signals are acquired with the corresponding time information,
it is possible to analyze data by selecting an appropriate time range after
the measurement.
For muonium atoms, this method, which is called the old muonium method,
was also applied to improve the measurement precision~\cite{PRA52.1948}.


The Rabi-oscillation spectroscopy presented here
is a novel technique where the detuning frequency
is directly obtained from the Rabi oscillation.
Let us consider the time evolution, known as Rabi oscillation, of transition signals from a two-state atom
interacting with a microwave field.
When the microwave frequency is detuned away from the resonance center,
the peak of the signal becomes lower, and the oscillations become faster.
Therefore, the detuning frequency can be obtained from
the time evolution of the Rabi oscillation if the microwave field strength is known.

  \begin{figure}
   \centering
   \includegraphics[width=8.6cm, bb= 1 1 589 497]{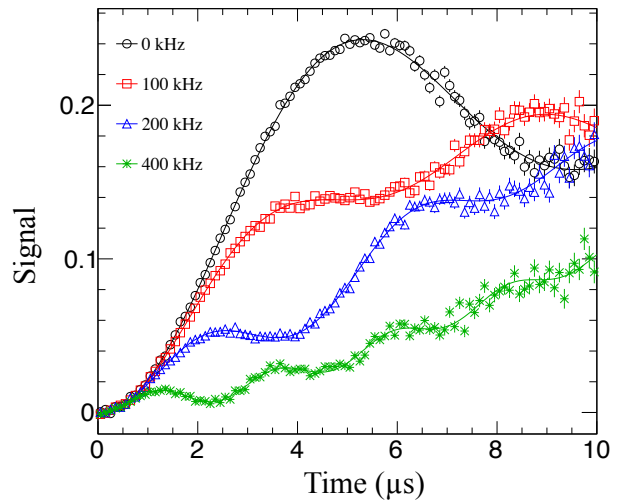}
   \caption{Simulation results of the time evolution of resonance signals
   for muonium atoms at different detuning frequencies of 0, 100, 200, and 400~kHz.
   Error bars represent the statistical uncertainties.}
   \label{225925_12May18}
  \end{figure}

In the case of the muonium HFS measurement, 
the time evolution of muon polarization can be obtained by measuring 
the change of the number of decay positrons
at a downstream or upstream counter because
the positron emission direction is preferentially along the muon spin direction.
The muon spin-flip signal due to the Rabi oscillation can be extracted
if we define this signal as the ratio of the number of 
positrons with the microwave present, $N_{\rm ON}$, and the number of 
positrons without the microwave, $N_{\rm OFF}$, minus one.
In a zero magnetic field, 
we do not measure the simple signal from a transition between two states,
but from singlet-triplet transitions among four states, resulting in a combination of two different oscillation terms.
The time evolution of the signal is described as
 \begin{eqnarray}
  {S \left( t \right)} &=& \frac{ N_{\rm ON}\left(t\right)}{ N_{\rm OFF}\left(t\right)}-1 \nonumber \\ 
   &=&
     A\ \left( \frac{G^{+}}{\Gamma}\cos G^{-}t
   + \frac{G^{-}}{\Gamma}\cos G^{+}t -1
       \right), \label{183706_15May18}
 \end{eqnarray}
 in which $A$ is a constant depending on the detector acceptance and
 energy threshold of decay positrons, and
 \begin{eqnarray}
  G^{\pm}&=&\frac{\Gamma\pm \Delta\omega}{2}, \\
  \Gamma&=&\sqrt{\left(\Delta\omega\right)^2+8\left|b\right|^2},\label{073603_8Apr20}
 \end{eqnarray}
  where $\Delta\omega$ is the detuning angular frequency and
  $|b|$ is a parameter proportional to the strength of the applied microwave field.
  The time dependence of the signal is expressed by
  the summation of cosine functions.
  Their oscillation frequencies and amplitudes are related to the detuning frequency and 
  the field strength of the stored microwave.
  When the microwave frequency is further detuned from the HFS resonance,
  the frequency $G^{-}$ decreases and $G^{+}$ increases,
  while the signal amplitude ${G^{+}}/{\Gamma}$ 
  increases and ${G^{-}}/{\Gamma}$ decreases.
  When the microwave power increases,
  both frequencies, $G^{-}$ and $G^{+}$, increase, while
  the signal amplitude ${G^{+}}/{\Gamma}$ 
  decreases and ${G^{-}}/{\Gamma}$ increases.
  The response to the signal form is different for the microwave
  frequency and the stored microwave power.
  Hence, the Rabi-oscillation analysis can extract the muonium HFS and
  the microwave power at the same time, both from the time evolution of the resonance signal.
  This also indicates that there is no need to sweep the microwave frequency.
  Therefore, the Rabi-oscillation analysis does not suffer from the uncertainty due to a systematic factor 
  that makes the resonance curve asymmetric.

  The signals observed in the measurement are, in fact, more complicated
  because of the spacial distribution of the muonium atoms and the variance of the magnetic field strength of the microwave.
  The distribution of the microwave power felt by muonium atoms was evaluated
  by referring to a calculated field map in the microwave cavity~\cite{Tanaka.PhD.Thesis} 
  and summing up over the muon stopping distribution estimated with a particle-tracking 
  simulation program based on Geant4 toolkit~\cite{AGOSTINELLI2003250, 1610988, ALLISON2016186}.
  Then, the time evolution of the resonance signal was calculated 
  by estimating the time distribution of the number of detected positrons in each case with and without microwaves.
  Statistical fluctuations were calculated according to the Poisson distribution.
  Simulation results of the time-evolution signals for several detuning frequencies 
  are plotted in Fig.~\ref{225925_12May18}.
  These simulated signals were then used as a data set for the Rabi-oscillation analysis
  to see if the correct detuning frequencies can be extracted back by the data fitting.
  The fitting function is expressed as
  \begin{widetext}
  \begin{eqnarray}
   f\left(t;\ A,\ |b|,\ \Delta\omega \right)
    &=& A\ \sum^{}_{i} N_{i}
    \left( \frac{G^{+}_{i}}{\Gamma_{i}}\cos G^{-}_{i}t
     + \frac{G^{-}_{i}}{\Gamma_{i}}\cos G^{+}_{i}t
     - 1 \right),
   \label{131155_17May18}
  \end{eqnarray}
  \end{widetext}
  where
  $i$ represents the index of summation over the discrete muon stopping position,
  and $N_{i}$ is the fraction of stopping muons at that position,
  with
  $A$, $|b|$ and $\Delta\omega$ being free parameters.

  It must be noted that the absolute value of
  the detuning frequency $\left|\Delta\omega\right|$
  can be obtained from the Rabi-oscillation analysis, but not its sign;
  the reason is that a negative detuning gives exactly the same Rabi oscillation
  as a positive detuning, as is evident from Eqs. (\ref{183706_15May18})-(\ref{073603_8Apr20}).
  The muonium HFS frequency $\nu_{\rm HFS}$ is obtained from the following equation:
  \begin{eqnarray}
   \left|\Delta\omega\right|/2\pi = \left| \nu_{\rm mw} - \nu_{\rm HFS} \right|,  \label{230053_14Mar18}
  \end{eqnarray}
  where $\nu_{\rm mw}$ is the microwave frequency.

  \begin{figure}
   \centering
   \includegraphics[width=8.6cm, bb=0 0 593 1185]{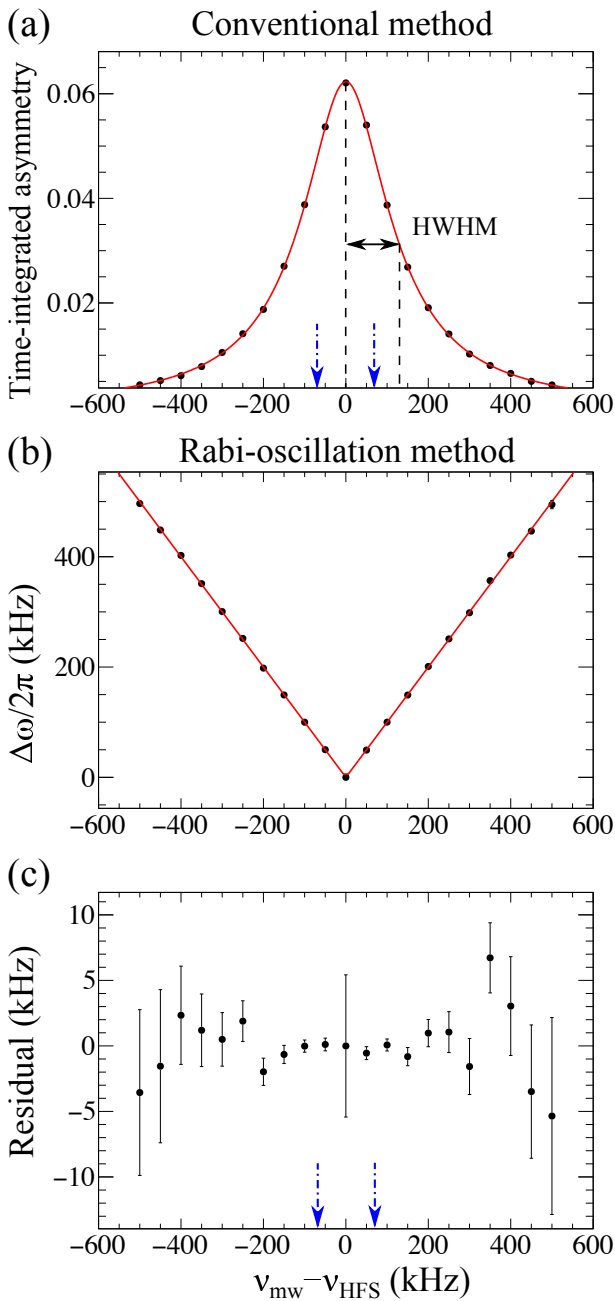}
   \caption{Simulation results of the conventional (a) and the Rabi-oscillation methods (b).
   The solid lines indicate the fitting results. The graph (c) shows the residuals of the Rabi-oscillation method.
   The black dashed lines and the horizontal arrow in the graph (a) indicate the HWHM of the resonance curve, 
   and the blue dash-dotted arrows in graphs (a) and (c) represent the detuning frequency with the lowest uncertainty in the
   Rabi-oscillation spectroscopy.
   }
   \label{170125_9Oct20}
  \end{figure}
  \begin{figure}
   \centering
   \includegraphics[width=8.6cm, bb=0 0 569 409]{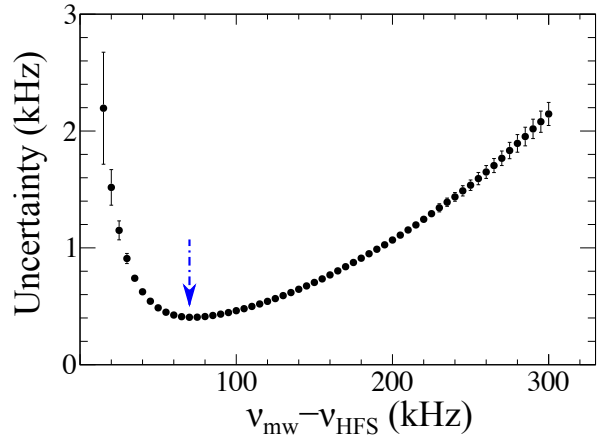}
   \caption{The detuning-frequency dependence of fitting uncertainties of the resonance frequency 
   obtained by repeating 500 times for each frequency the same simulation of 
   the Rabi-oscillation spectroscopy as in Fig.~\ref{170125_9Oct20}. 
   The error bars represent standard deviations.
   }
   \label{134608_25Nov20}
  \end{figure}

  The results of simulated measurements at various microwave frequencies are shown in Fig.~\ref{170125_9Oct20}.
  The resonance frequency can be obtained by both the conventional and the Rabi-oscillation methods.
  The fitting results of the simulation assure that
  we can successfully obtain the resonance frequency
  from the Rabi-oscillation analysis for different microwave frequencies.
  The plot of the residuals in Fig.~\ref{170125_9Oct20}\,(c) shows that
  the uncertainty depends on the detuning frequency in the Rabi-oscillation spectroscopy.
  The uncertainty in determining the detuning frequency becomes large when the microwave is tuned on resonance,
  because fitting parameters $|b|$ and $\Delta\omega$ are correlated and difficult to separate.
  The uncertainty is also large when the microwave frequency is far away from the resonance, 
  because most muons decay before the spin-flip due to the slow Rabi oscillation.
  The same simulation as in Fig.~\ref{170125_9Oct20} was repeated 500 times for each frequency 
  to investigate the detuning-frequency dependence of the uncertainty in determining the resonance frequency 
  by the Rabi-oscillation method, and the results are shown in Fig.~\ref{134608_25Nov20}.
  The best precision is achieved for a detuning frequency
  of about $\pm 70$\,kHz, 
  corresponding to half of the HWHM (half width at half maximum) of the resonance curve in the frequency domain.
  According to our simulation study, the Rabi-oscillation method can improve precision nearly twice as much as the conventional method
  by concentrating all data at a 
  detuning frequency of about $\pm 70$\,kHz.

  \begin{figure*}
   \centering
   \includegraphics[width=17.2cm, bb=1 1 598 352]
   {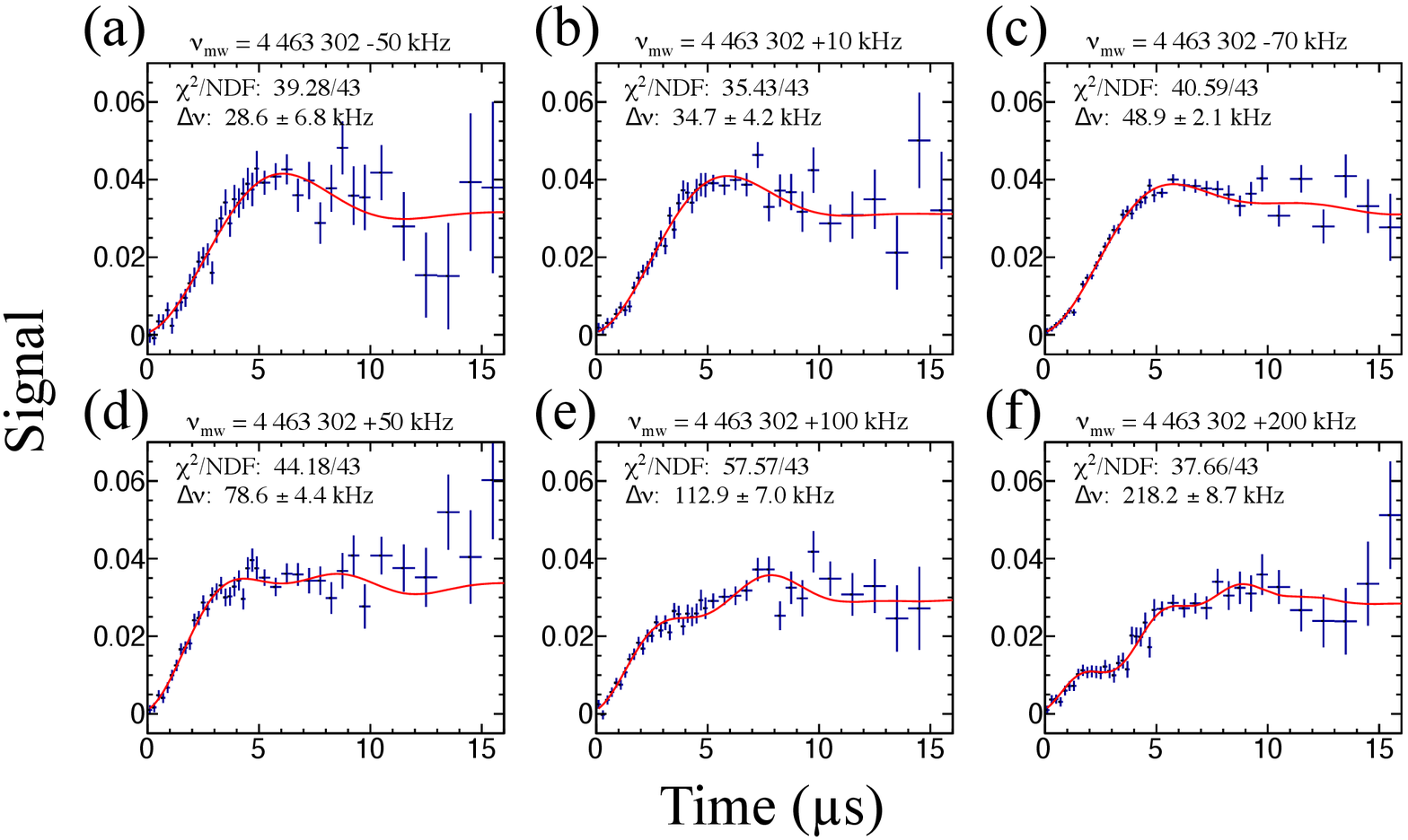}
   \caption{Experimental results with the Rabi-oscillation spectroscopy.
   Graphs (a)-(f) show the time evolution of resonance signals with microwave frequencies of 4~463~302~kHz $-$50, +10,
   $-$70, +50, +100, and +200~kHz, respectively, and are shown in ascending order of the absolute value of the detuning frequency.
   The solid lines indicate the fitting results. }
   \label{184838_31Oct19}
  \end{figure*}

There were experiments in the past that tried to optionally use time-dependent information~\cite{PhysRevA.5.2338,PhysRevLett.23.513,PhysRevLett.27.474},
in order to confirm the results of the conventional method. 
The Rabi-oscillation spectroscopy, which makes full use of time information, requires high statistics and a good time-accurate detector,
which was not technically possible in the past.

New precise measurements of the muonium ground state HFS 
are now in progress using the high-intensity pulsed muon beam 
at the Japan Proton Accelerator Research Complex (J-PARC)
by the Muonium Spectroscopy Experiment Using Microwave (MuSEUM) collaboration~\cite{doi:10.1063/1.3644324}.
In our recent experiment~\cite{k2020new},
a polarized muon beam was injected into a krypton gas target,
resulting in the formation of muonium.
The muon spin-flip signal induced by the microwave was measured with
downstream positron detectors
that use a silicon strip sensor~\cite{Aoyagi_2020}.
The sensor was optimized to the pulsed beam experiment and
exhibited a high-rate capability to reduce inefficiency due to pile-up events.
The use of a high intensity beam and a detector with a good time response, as described above,
were important keys in the technical development needed to make
the Rabi-oscillation spectroscopy successful for the first time.

\begin{figure}
   \centering
   \includegraphics[width=8.6cm, bb=12 8 595 413]
   {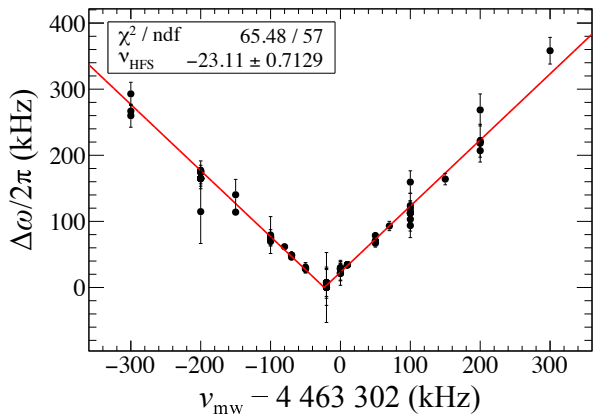}
   \caption{Determination of $\nu_{\rm HFS}$ from the fitting results of
   the Rabi-oscillation analysis for different microwave
   frequencies. The data were for a krypton gas target 
   with a pressure of about $7\times10^{4}$~Pa.
   }
   \label{182052_5Nov19}
  \end{figure}

  We have analyzed a part of the experimental data obtained in June 2018 using the Rabi-oscillation
  method. The target gas was krypton
  with a pressure of $7.003\times10^{4}$~Pa at room temperature.
  The measurement time at that condition was about 42~hours.
  The results of the Rabi-oscillation analysis are shown in
  Fig.~\ref{184838_31Oct19}.
  The time of the muon injection was calibrated by a time histogram
  of detected positrons without the microwave.
  We obtained a multitude of data sets with different frequencies,
  and applied the Rabi-oscillation analysis to extract the muonium HFS frequency.
  The results plotted in Fig.~\ref{182052_5Nov19}
  show the microwave frequency in the horizontal axis and
  the detuning frequency obtained by the Rabi-oscillation analysis
  in the vertical axis.
The data contain measurements at different microwave powers, and yet the data points are well fitted together in a straight line,
revealing the advantageous features of the Rabi-oscillation spectroscopy which can make use of all the data
available regardless of difference in the microwave input power.
  
\begin{table}[h] 
 \caption{Summary of systematic uncertainties}
 \label{142236_9Jun20}
  \begin{ruledtabular}
  \begin{tabular}{ccc}
   Contribution & June 2018 & Prospects \\ \hline
   Precision of pressure gauge & 46 & 5\\
   Pressure fluctuation & 25 & 5\\
   Precision of microwave frequency & 45 & $<$ 1\\
   Microwave power drift & 10 & $<$ 1 \\
   Detector pileup & 1 & $<$ 1 \\
   Time accuracy of detector & 1 & 1
  \end{tabular}
 \end{ruledtabular}
\end{table}

  By correcting for the gas density shift of the resonance frequency
  due to atomic collision~\cite{PhysRevA.8.86}, 
  the extrapolated resonance frequency at zero density 
  $\left(\rho = 0\right)$ was determined as
  \begin{eqnarray}
   \nu_{\rm HFS }\left(0\right)
 = \left( 4\ 463\ 301.61 \pm 0.71 \right) \ {\rm kHz}\ 
 \left(\rm 160\ ppb\right),
  \end{eqnarray}
  where the uncertainty shown is of a statistical nature.

  Dominant systematic uncertainties
  are listed in Table~\ref{142236_9Jun20}.
  The systematic uncertainties consist of 
  the uncertainty due to the precision of the gas pressure determination, the uncertainty due to the microwaves in the cavity, and the uncertainty due to the performance of the detector. 
  The precision of the gas pressure determination was caused by the precision of the pressure gauge and the pressure fluctuation due to the gas temperature change, 
  and can be improved up to 5~Hz by using a pressure gauge with a precision of 0.02$\%$
  and a temperature control device, respectively. 
  One of the microwave-induced uncertainties came from the lack of accuracy of the microwave frequency reference, 
  which can be reduced to less than 1~Hz by using GPS clock system as a reference. 
  The other was the effect of the cavity warming up and changing the Q-value during the same frequency measurement, 
  which can be improved to less than 1~Hz by using a water cooling system.
   A detailed description will be given in a separate paper.
   
   The obtained result is consistent with the previous two experiments 
   at Los Alamos Meson Physics Facility~\cite{CASPERSON1975397, PhysRevLett.82.711}
   within two standard deviations, and 1.8 times as precise as
   the previous experiment at the zero magnetic field~\cite{CASPERSON1975397,PhysRevA.8.86}.
  
 Thus, we have successfully established a new spectroscopic method, which we named Rabi-oscillation spectroscopy, 
 and have obtained the resonance frequency directly from the time evolution of the Rabi oscillation without drawing a resonance curve, 
 eliminating the need for a frequency sweep. 
 By using a fitting function that can treat the measurement data with different conditions in a unified manner, 
 we have also established a method for determining the resonance frequency stably from data under various experimental conditions, 
 which can not be handled together by conventional methods.

  Rabi-oscillation spectroscopy is applicable to the measurement of resonance frequencies of atoms and molecules containing unstable particles, 
  especially those produced in limited quantities at accelerators, such as exotic atoms and isotopes with short-lived nuclei.
  The beam time is finite and this technique is clearly effective as it does not require adjustment of microwave power or frequency sweeping.

  This work was conducted under a user program
  Proposal No. 2017A0134, 2017B0224, and 2018A0071 at J-PARC MLF.
  The authors would like to acknowledge the help and expertise
  of the staff of J-PARC MLF MUSE.
  This work was supported by Japanese JSPS KAKENHI Grant Number
  JP23244046,
  JP26247046,
  JP15H05742,
  JP17H01133,
  and
  JP19K14746.
\bibliography{string_abbrv, ref}

\begin{thebibliography}{18}%
\makeatletter
\providecommand \@ifxundefined [1]{%
 \@ifx{#1\undefined}
}%
\providecommand \@ifnum [1]{%
 \ifnum #1\expandafter \@firstoftwo
 \else \expandafter \@secondoftwo
 \fi
}%
\providecommand \@ifx [1]{%
 \ifx #1\expandafter \@firstoftwo
 \else \expandafter \@secondoftwo
 \fi
}%
\providecommand \natexlab [1]{#1}%
\providecommand \enquote  [1]{``#1''}%
\providecommand \bibnamefont  [1]{#1}%
\providecommand \bibfnamefont [1]{#1}%
\providecommand \citenamefont [1]{#1}%
\providecommand \href@noop [0]{\@secondoftwo}%
\providecommand \href [0]{\begingroup \@sanitize@url \@href}%
\providecommand \@href[1]{\@@startlink{#1}\@@href}%
\providecommand \@@href[1]{\endgroup#1\@@endlink}%
\providecommand \@sanitize@url [0]{\catcode `\\12\catcode `\$12\catcode
  `\&12\catcode `\#12\catcode `\^12\catcode `\_12\catcode `\%12\relax}%
\providecommand \@@startlink[1]{}%
\providecommand \@@endlink[0]{}%
\providecommand \url  [0]{\begingroup\@sanitize@url \@url }%
\providecommand \@url [1]{\endgroup\@href {#1}{\urlprefix }}%
\providecommand \urlprefix  [0]{URL }%
\providecommand \Eprint [0]{\href }%
\providecommand \doibase [0]{https://doi.org/}%
\providecommand \selectlanguage [0]{\@gobble}%
\providecommand \bibinfo  [0]{\@secondoftwo}%
\providecommand \bibfield  [0]{\@secondoftwo}%
\providecommand \translation [1]{[#1]}%
\providecommand \BibitemOpen [0]{}%
\providecommand \bibitemStop [0]{}%
\providecommand \bibitemNoStop [0]{.\EOS\space}%
\providecommand \EOS [0]{\spacefactor3000\relax}%
\providecommand \BibitemShut  [1]{\csname bibitem#1\endcsname}%
\let\auto@bib@innerbib\@empty
\bibitem [{\citenamefont {Mohr}\ \emph {et~al.}(2016)\citenamefont {Mohr},
  \citenamefont {Newell},\ and\ \citenamefont {Taylor}}]{RevModPhys.88.035009}%
  \BibitemOpen
  \bibfield  {author} {\bibinfo {author} {\bibfnamefont {P.~J.}\ \bibnamefont
  {Mohr}}, \bibinfo {author} {\bibfnamefont {D.~B.}\ \bibnamefont {Newell}},\
  and\ \bibinfo {author} {\bibfnamefont {B.~N.}\ \bibnamefont {Taylor}},\
  }\bibfield  {title} {\bibinfo {title} {{CODATA} recommended values of the
  fundamental physical constants: 2014},\ }\href
  {https://doi.org/10.1103/RevModPhys.88.035009} {\bibfield  {journal}
  {\bibinfo  {journal} {Rev. Mod. Phys.}\ }\textbf {\bibinfo {volume} {88}},\
  \bibinfo {pages} {035009} (\bibinfo {year} {2016})}\BibitemShut {NoStop}%
\bibitem [{\citenamefont {Keshavarzi}\ \emph {et~al.}(2020)\citenamefont
  {Keshavarzi}, \citenamefont {Nomura},\ and\ \citenamefont
  {Teubner}}]{PhysRevD.101.014029}%
  \BibitemOpen
  \bibfield  {author} {\bibinfo {author} {\bibfnamefont {A.}~\bibnamefont
  {Keshavarzi}}, \bibinfo {author} {\bibfnamefont {D.}~\bibnamefont {Nomura}},\
  and\ \bibinfo {author} {\bibfnamefont {T.}~\bibnamefont {Teubner}},\
  }\bibfield  {title} {\bibinfo {title} {$g\ensuremath{-}2$ of charged leptons,
  $\ensuremath{\alpha}\mathbf{(}{M}_{Z}^{2}\mathbf{)}$, and the hyperfine
  splitting of muonium},\ }\href {https://doi.org/10.1103/PhysRevD.101.014029}
  {\bibfield  {journal} {\bibinfo  {journal} {Phys. Rev. D}\ }\textbf {\bibinfo
  {volume} {101}},\ \bibinfo {pages} {014029} (\bibinfo {year}
  {2020})}\BibitemShut {NoStop}%
\bibitem [{\citenamefont {Abe}\ \emph {et~al.}(2019)\citenamefont {Abe},
  \citenamefont {Bae}, \citenamefont {Beer}, \citenamefont {Bunce},
  \citenamefont {Choi}, \citenamefont {Choi}, \citenamefont {Chung},
  \citenamefont {da~Silva}, \citenamefont {Eidelman}, \citenamefont {Finger},
  \citenamefont {Fukao}, \citenamefont {Fukuyama}, \citenamefont
  {Haciomeroglu}, \citenamefont {Hasegawa}, \citenamefont {Hayasaka},
  \citenamefont {Hayashizaki}, \citenamefont {Hisamatsu}, \citenamefont
  {Iijima}, \citenamefont {Iinuma}, \citenamefont {Ikeda}, \citenamefont
  {Ikeno}, \citenamefont {Inami}, \citenamefont {Ishida}, \citenamefont
  {Itahashi}, \citenamefont {Iwasaki}, \citenamefont {Iwashita}, \citenamefont
  {Iwata}, \citenamefont {Kadono}, \citenamefont {Kamal}, \citenamefont
  {Kamitani}, \citenamefont {Kanda}, \citenamefont {Kapusta}, \citenamefont
  {Kawagoe}, \citenamefont {Kawamura}, \citenamefont {Kim}, \citenamefont
  {Kim}, \citenamefont {Kishishita}, \citenamefont {Kitamura}, \citenamefont
  {Ko}, \citenamefont {Kohriki}, \citenamefont {Kondo}, \citenamefont {Kume},
  \citenamefont {Lee}, \citenamefont {Lee}, \citenamefont {Lee}, \citenamefont
  {Marshall}, \citenamefont {Matsuda}, \citenamefont {Mibe}, \citenamefont
  {Miyake}, \citenamefont {Murakami}, \citenamefont {Nagamine}, \citenamefont
  {Nakayama}, \citenamefont {Nishimura}, \citenamefont {Nomura}, \citenamefont
  {Ogitsu}, \citenamefont {Ohsawa}, \citenamefont {Oide}, \citenamefont
  {Oishi}, \citenamefont {Okada}, \citenamefont {Olin}, \citenamefont {Omarov},
  \citenamefont {Otani}, \citenamefont {Razuvaev}, \citenamefont {Rehman},
  \citenamefont {Saito}, \citenamefont {Saito}, \citenamefont {Sasaki},
  \citenamefont {Sasaki}, \citenamefont {Sato}, \citenamefont {Sato},
  \citenamefont {Semertzidis}, \citenamefont {Sendai}, \citenamefont
  {Shatunov}, \citenamefont {Shimomura}, \citenamefont {Shoji}, \citenamefont
  {Shwartz}, \citenamefont {Strasser}, \citenamefont {Sue}, \citenamefont
  {Suehara}, \citenamefont {Sung}, \citenamefont {Suzuki}, \citenamefont
  {Takatomi}, \citenamefont {Tanaka}, \citenamefont {Tojo}, \citenamefont
  {Tsutsumi}, \citenamefont {Uchida}, \citenamefont {Ueno}, \citenamefont
  {Wada}, \citenamefont {Won}, \citenamefont {Yamaguchi}, \citenamefont
  {Yamanaka}, \citenamefont {Yamamoto}, \citenamefont {Yamazaki}, \citenamefont
  {Yasuda}, \citenamefont {Yoshida},\ and\ \citenamefont
  {Yoshioka}}]{10.1093ptepptz030}%
  \BibitemOpen
  \bibfield  {author} {\bibinfo {author} {\bibfnamefont {M.}~\bibnamefont
  {Abe}} \emph {et~al.},\ }\bibfield  {title} {\bibinfo {title} {{A new
  approach for measuring the muon anomalous magnetic moment and electric dipole
  moment}},\ }\href@noop {} {\bibfield  {journal} {\bibinfo  {journal} {Prog.
  Theor. Exp. Phys.}\ }\textbf {\bibinfo {volume} {2019}} (\bibinfo {year}
  {2019})},\ \bibinfo {note} {053C02}\BibitemShut {NoStop}%
\bibitem [{\citenamefont {Grange}\ \emph {et~al.}(2015)\citenamefont {Grange},
  \citenamefont {Guarino}, \citenamefont {Winter}, \citenamefont {Wood},
  \citenamefont {Zhao}, \citenamefont {Carey}, \citenamefont {Gastler},
  \citenamefont {Hazen}, \citenamefont {Kinnaird}, \citenamefont {Miller},
  \citenamefont {Mott}, \citenamefont {Roberts}, \citenamefont {Benante},
  \citenamefont {Crnkovic}, \citenamefont {Morse}, \citenamefont {Sayed},
  \citenamefont {Tishchenko}, \citenamefont {Druzhinin}, \citenamefont
  {Khazin}, \citenamefont {Koop}, \citenamefont {Logashenko}, \citenamefont
  {Shatunov}, \citenamefont {Solodov}, \citenamefont {Korostelev},
  \citenamefont {Newton}, \citenamefont {Wolski}, \citenamefont {Chapelain},
  \citenamefont {Bjorkquist}, \citenamefont {Eggert}, \citenamefont
  {Frankenthal}, \citenamefont {Gibbons}, \citenamefont {Kim}, \citenamefont
  {Mikhailichenko}, \citenamefont {Orlov}, \citenamefont {Rubin}, \citenamefont
  {Sweigart}, \citenamefont {Allspach}, \citenamefont {Annala}, \citenamefont
  {Barzi}, \citenamefont {Bourland}, \citenamefont {Brown}, \citenamefont
  {Casey}, \citenamefont {Chappa}, \citenamefont {Convery}, \citenamefont
  {Drendel}, \citenamefont {Friedsam}, \citenamefont {Gadfort}, \citenamefont
  {Hardin}, \citenamefont {Hawke}, \citenamefont {Hayes}, \citenamefont
  {Jaskierny}, \citenamefont {Johnstone}, \citenamefont {Johnstone},
  \citenamefont {Kashikhin}, \citenamefont {Kendziora}, \citenamefont {Kiburg},
  \citenamefont {Klebaner}, \citenamefont {Kourbanis}, \citenamefont {Kyle},
  \citenamefont {Larson}, \citenamefont {Leveling}, \citenamefont {Lyon},
  \citenamefont {Markley}, \citenamefont {McArthur}, \citenamefont {Merritt},
  \citenamefont {Mokhov}, \citenamefont {Morgan}, \citenamefont {Nguyen},
  \citenamefont {Ostiguy}, \citenamefont {Para}, \citenamefont {Popovic},
  \citenamefont {Ramberg}, \citenamefont {Rominsky}, \citenamefont {Schoo},
  \citenamefont {Schultz}, \citenamefont {Still}, \citenamefont {Soha},
  \citenamefont {Strigonov}, \citenamefont {Tassotto}, \citenamefont
  {Turrioni}, \citenamefont {Villegas}, \citenamefont {Voirin}, \citenamefont
  {Velev}, \citenamefont {Welty-Rieger}, \citenamefont {Wolff}, \citenamefont
  {Worel}, \citenamefont {Wu}, \citenamefont {Zifko}, \citenamefont {Jungmann},
  \citenamefont {Onderwater}, \citenamefont {Debevec}, \citenamefont {Ganguly},
  \citenamefont {Kasten}, \citenamefont {Leo}, \citenamefont {Pitts},
  \citenamefont {Schlesier}, \citenamefont {Gaisser}, \citenamefont
  {Haciomeroglu}, \citenamefont {Kim}, \citenamefont {Lee}, \citenamefont
  {Lee}, \citenamefont {Semertzidis}, \citenamefont {Giovanetti}, \citenamefont
  {Baranov}, \citenamefont {Duginov}, \citenamefont {Khomutov}, \citenamefont
  {Krylov}, \citenamefont {Kuchinskiy}, \citenamefont {Volnykh}, \citenamefont
  {Crawford}, \citenamefont {Fatemi}, \citenamefont {Gohn}, \citenamefont
  {Gorringe}, \citenamefont {Korsch}, \citenamefont {Plaster}, \citenamefont
  {Anastasi}, \citenamefont {Babusci}, \citenamefont {Dabagov}, \citenamefont
  {Ferrari}, \citenamefont {Fioretti}, \citenamefont {Gabbanini}, \citenamefont
  {Hampai}, \citenamefont {Palladino}, \citenamefont {Venanzoni}, \citenamefont
  {Bowcock}, \citenamefont {Carroll}, \citenamefont {King}, \citenamefont
  {Maxfield}, \citenamefont {McCormick}, \citenamefont {Price}, \citenamefont
  {Sim}, \citenamefont {Smith}, \citenamefont {Teubner}, \citenamefont
  {Turner}, \citenamefont {Whitley}, \citenamefont {Wormald}, \citenamefont
  {Chislett}, \citenamefont {Kilani}, \citenamefont {Lancaster}, \citenamefont
  {Motuk}, \citenamefont {Stuttard}, \citenamefont {Warren}, \citenamefont
  {Flay}, \citenamefont {Kawall}, \citenamefont {Meadows}, \citenamefont
  {Chupp}, \citenamefont {Raymond}, \citenamefont {Tewlsey-Booth},
  \citenamefont {Syphers}, \citenamefont {Tarazona}, \citenamefont
  {Catalonotti}, \citenamefont {Stefano}, \citenamefont {Iacovacci},
  \citenamefont {Mastroianni}, \citenamefont {Chattopadhyay}, \citenamefont
  {Eads}, \citenamefont {Fortner}, \citenamefont {Hedin}, \citenamefont
  {Pohlman}, \citenamefont {de~Gouvea}, \citenamefont {Schellman},
  \citenamefont {Welty-Rieger}, \citenamefont {Azfar}, \citenamefont {Henry},
  \citenamefont {Alkhazov}, \citenamefont {Golovtsov}, \citenamefont
  {Neustroev}, \citenamefont {Uvarov}, \citenamefont {Vasilyev}, \citenamefont
  {Vorobyov}, \citenamefont {Zhalov}, \citenamefont {Cerrito}, \citenamefont
  {Gray}, \citenamefont {Sciascio}, \citenamefont {Moricciani}, \citenamefont
  {Fu}, \citenamefont {Ji}, \citenamefont {Li}, \citenamefont {Yang},
  \citenamefont {Stöckinger}, \citenamefont {Cantatore}, \citenamefont {Cauz},
  \citenamefont {Karuza}, \citenamefont {Pauletta}, \citenamefont {Santi},
  \citenamefont {Baeßler}, \citenamefont {Bychkov}, \citenamefont {Frlez},
  \citenamefont {Pocanic}, \citenamefont {Alonzi}, \citenamefont {Fertl},
  \citenamefont {Fienberg}, \citenamefont {Froemming}, \citenamefont {Garcia},
  \citenamefont {Kaspar}, \citenamefont {Kammel}, \citenamefont {Osofsky},
  \citenamefont {Smith}, \citenamefont {Swanson}, \citenamefont {van Wechel},\
  and\ \citenamefont {Lynch}}]{grange2015muon}%
  \BibitemOpen
  \bibfield  {author} {\bibinfo {author} {\bibfnamefont {J.}~\bibnamefont
  {Grange}} \emph {et~al.},\ }\href@noop {} {\bibinfo {title} {Muon ($g-2$)
  technical design report}} (\bibinfo {year} {2015}),\ \Eprint
  {https://arxiv.org/abs/1501.06858} {arXiv:1501.06858 [physics.ins-det]}
  \BibitemShut {NoStop}%
\bibitem [{\citenamefont {Thompson}\ \emph {et~al.}(1973)\citenamefont
  {Thompson}, \citenamefont {Crane}, \citenamefont {Crane}, \citenamefont
  {Amato}, \citenamefont {Hughes}, \citenamefont {Putlitz},\ and\ \citenamefont
  {Rothberg}}]{PhysRevA.8.86}%
  \BibitemOpen
  \bibfield  {author} {\bibinfo {author} {\bibfnamefont {P.~A.}\ \bibnamefont
  {Thompson}}, \bibinfo {author} {\bibfnamefont {P.}~\bibnamefont {Crane}},
  \bibinfo {author} {\bibfnamefont {T.}~\bibnamefont {Crane}}, \bibinfo
  {author} {\bibfnamefont {J.~J.}\ \bibnamefont {Amato}}, \bibinfo {author}
  {\bibfnamefont {V.~W.}\ \bibnamefont {Hughes}}, \bibinfo {author}
  {\bibfnamefont {G.~z.}\ \bibnamefont {Putlitz}},\ and\ \bibinfo {author}
  {\bibfnamefont {J.~E.}\ \bibnamefont {Rothberg}},\ }\bibfield  {title}
  {\bibinfo {title} {Muonium. {IV}. precision measurement of the muonium
  hyperfine-structure interval at weak and very weak magnetic fields},\ }\href
  {https://doi.org/10.1103/PhysRevA.8.86} {\bibfield  {journal} {\bibinfo
  {journal} {Phys. Rev. A}\ }\textbf {\bibinfo {volume} {8}},\ \bibinfo {pages}
  {86} (\bibinfo {year} {1973})}\BibitemShut {NoStop}%
\bibitem [{\citenamefont {Casperson}\ \emph {et~al.}(1975)\citenamefont
  {Casperson}, \citenamefont {Crane}, \citenamefont {Hughes}, \citenamefont
  {Souder}, \citenamefont {Stambaugh}, \citenamefont {Thompson}, \citenamefont
  {Orth}, \citenamefont {zu~Putlitz}, \citenamefont {Kaspar}, \citenamefont
  {Reist},\ and\ \citenamefont {Denison}}]{CASPERSON1975397}%
  \BibitemOpen
  \bibfield  {author} {\bibinfo {author} {\bibfnamefont {D.}~\bibnamefont
  {Casperson}} \emph {et~al.},\ }\bibfield  {title} {\bibinfo {title} {A new
  high precision measurement of the muonium hyperfine structure interval
  $\delta \nu$},\ }\href
  {https://doi.org/https://doi.org/10.1016/0370-2693(75)90099-4} {\bibfield
  {journal} {\bibinfo  {journal} {Phys. Lett. B}\ }\textbf {\bibinfo {volume}
  {59}},\ \bibinfo {pages} {397 } (\bibinfo {year} {1975})}\BibitemShut
  {NoStop}%
\bibitem [{\citenamefont {Liu}\ \emph {et~al.}(1999)\citenamefont {Liu},
  \citenamefont {Boshier}, \citenamefont {Dhawan}, \citenamefont {van Dyck},
  \citenamefont {Egan}, \citenamefont {Fei}, \citenamefont {Grosse~Perdekamp},
  \citenamefont {Hughes}, \citenamefont {Janousch}, \citenamefont {Jungmann},
  \citenamefont {Kawall}, \citenamefont {Mariam}, \citenamefont {Pillai},
  \citenamefont {Prigl}, \citenamefont {zu~Putlitz}, \citenamefont {Reinhard},
  \citenamefont {Schwarz}, \citenamefont {Thompson},\ and\ \citenamefont
  {Woodle}}]{PhysRevLett.82.711}%
  \BibitemOpen
  \bibfield  {author} {\bibinfo {author} {\bibfnamefont {W.}~\bibnamefont
  {Liu}} \emph {et~al.},\ }\bibfield  {title} {\bibinfo {title} {High precision
  measurements of the ground state hyperfine structure interval of muonium and
  of the muon magnetic moment},\ }\href
  {https://doi.org/10.1103/PhysRevLett.82.711} {\bibfield  {journal} {\bibinfo
  {journal} {Phys. Rev. Lett.}\ }\textbf {\bibinfo {volume} {82}},\ \bibinfo
  {pages} {711} (\bibinfo {year} {1999})}\BibitemShut {NoStop}%
\bibitem [{\citenamefont {Boshier}\ \emph {et~al.}(1995)\citenamefont
  {Boshier}, \citenamefont {Dhawan}, \citenamefont {Fei}, \citenamefont
  {Hughes}, \citenamefont {Janousch}, \citenamefont {Jungmann}, \citenamefont
  {Liu}, \citenamefont {Pillai}, \citenamefont {Prigl}, \citenamefont
  {Putlitz}, \citenamefont {Reinhard}, \citenamefont {Schwarz}, \citenamefont
  {Souder}, \citenamefont {O}, \citenamefont {Wang}, \citenamefont {Woodle},\
  and\ \citenamefont {Xu}}]{PRA52.1948}%
  \BibitemOpen
  \bibfield  {author} {\bibinfo {author} {\bibfnamefont {M.}~\bibnamefont
  {Boshier}} \emph {et~al.},\ }\bibfield  {title} {\bibinfo {title}
  {Observation of resonance line narrowing for old muonium},\ }\href
  {https://doi.org/10.1103/PhysRevA.52.1948} {\bibfield  {journal} {\bibinfo
  {journal} {Phys. Rev. A}\ }\textbf {\bibinfo {volume} {52}},\ \bibinfo
  {pages} {1948} (\bibinfo {year} {1995})}\BibitemShut {NoStop}%
\bibitem [{\citenamefont {Tanaka}(2015)}]{Tanaka.PhD.Thesis}%
  \BibitemOpen
  \bibfield  {author} {\bibinfo {author} {\bibfnamefont {K.~S.}\ \bibnamefont
  {Tanaka}},\ }\emph {\bibinfo {title} {Measurement of muonium hyperfine
  structure at {J-PARC}}},\ \href@noop {} {Ph.D. thesis},\ \bibinfo  {school}
  {The University of Tokyo} (\bibinfo {year} {2015})\BibitemShut {NoStop}%
\bibitem [{\citenamefont {Agostinelli}\ \emph {et~al.}(2003)\citenamefont
  {Agostinelli}, \citenamefont {Allison}, \citenamefont {Amako}, \citenamefont
  {Apostolakis}, \citenamefont {Araujo}, \citenamefont {Arce}, \citenamefont
  {Asai}, \citenamefont {Axen}, \citenamefont {Banerjee}, \citenamefont
  {Barrand}, \citenamefont {Behner}, \citenamefont {Bellagamba}, \citenamefont
  {Boudreau}, \citenamefont {Broglia}, \citenamefont {Brunengo}, \citenamefont
  {Burkhardt}, \citenamefont {Chauvie}, \citenamefont {Chuma}, \citenamefont
  {Chytracek}, \citenamefont {Cooperman}, \citenamefont {Cosmo}, \citenamefont
  {Degtyarenko}, \citenamefont {Dell'Acqua}, \citenamefont {Depaola},
  \citenamefont {Dietrich}, \citenamefont {Enami}, \citenamefont {Feliciello},
  \citenamefont {Ferguson}, \citenamefont {Fesefeldt}, \citenamefont {Folger},
  \citenamefont {Foppiano}, \citenamefont {Forti}, \citenamefont {Garelli},
  \citenamefont {Giani}, \citenamefont {Giannitrapani}, \citenamefont {Gibin},
  \citenamefont {Cadenas]}, \citenamefont {González}, \citenamefont {Abril]},
  \citenamefont {Greeniaus}, \citenamefont {Greiner}, \citenamefont {Grichine},
  \citenamefont {Grossheim}, \citenamefont {Guatelli}, \citenamefont
  {Gumplinger}, \citenamefont {Hamatsu}, \citenamefont {Hashimoto},
  \citenamefont {Hasui}, \citenamefont {Heikkinen}, \citenamefont {Howard},
  \citenamefont {Ivanchenko}, \citenamefont {Johnson}, \citenamefont {Jones},
  \citenamefont {Kallenbach}, \citenamefont {Kanaya}, \citenamefont {Kawabata},
  \citenamefont {Kawabata}, \citenamefont {Kawaguti}, \citenamefont {Kelner},
  \citenamefont {Kent}, \citenamefont {Kimura}, \citenamefont {Kodama},
  \citenamefont {Kokoulin}, \citenamefont {Kossov}, \citenamefont {Kurashige},
  \citenamefont {Lamanna}, \citenamefont {Lampén}, \citenamefont {Lara},
  \citenamefont {Lefebure}, \citenamefont {Lei}, \citenamefont {Liendl},
  \citenamefont {Lockman}, \citenamefont {Longo}, \citenamefont {Magni},
  \citenamefont {Maire}, \citenamefont {Medernach}, \citenamefont {Minamimoto},
  \citenamefont {de~Freitas]}, \citenamefont {Morita}, \citenamefont
  {Murakami}, \citenamefont {Nagamatu}, \citenamefont {Nartallo}, \citenamefont
  {Nieminen}, \citenamefont {Nishimura}, \citenamefont {Ohtsubo}, \citenamefont
  {Okamura}, \citenamefont {O'Neale}, \citenamefont {Oohata}, \citenamefont
  {Paech}, \citenamefont {Perl}, \citenamefont {Pfeiffer}, \citenamefont {Pia},
  \citenamefont {Ranjard}, \citenamefont {Rybin}, \citenamefont {Sadilov},
  \citenamefont {Salvo]}, \citenamefont {Santin}, \citenamefont {Sasaki},
  \citenamefont {Savvas}, \citenamefont {Sawada}, \citenamefont {Scherer},
  \citenamefont {Sei}, \citenamefont {Sirotenko}, \citenamefont {Smith},
  \citenamefont {Starkov}, \citenamefont {Stoecker}, \citenamefont {Sulkimo},
  \citenamefont {Takahata}, \citenamefont {Tanaka}, \citenamefont {Tcherniaev},
  \citenamefont {Tehrani]}, \citenamefont {Tropeano}, \citenamefont {Truscott},
  \citenamefont {Uno}, \citenamefont {Urban}, \citenamefont {Urban},
  \citenamefont {Verderi}, \citenamefont {Walkden}, \citenamefont {Wander},
  \citenamefont {Weber}, \citenamefont {Wellisch}, \citenamefont {Wenaus},
  \citenamefont {Williams}, \citenamefont {Wright}, \citenamefont {Yamada},
  \citenamefont {Yoshida},\ and\ \citenamefont
  {Zschiesche}}]{AGOSTINELLI2003250}%
  \BibitemOpen
  \bibfield  {author} {\bibinfo {author} {\bibfnamefont {S.}~\bibnamefont
  {Agostinelli}} \emph {et~al.},\ }\bibfield  {title} {\bibinfo {title}
  {Geant4—a simulation toolkit},\ }\href
  {https://doi.org/https://doi.org/10.1016/S0168-9002(03)01368-8} {\bibfield
  {journal} {\bibinfo  {journal} {Nucl. Inst. \& Meth. in Phys. Res. A}\
  }\textbf {\bibinfo {volume} {506}},\ \bibinfo {pages} {250 } (\bibinfo {year}
  {2003})}\BibitemShut {NoStop}%
\bibitem [{\citenamefont {{Allison}}\ \emph {et~al.}(2006)\citenamefont
  {{Allison}}, \citenamefont {{Amako}}, \citenamefont {{Apostolakis}},
  \citenamefont {{Araujo}}, \citenamefont {{Arce Dubois}}, \citenamefont
  {{Asai}}, \citenamefont {{Barrand}}, \citenamefont {{Capra}}, \citenamefont
  {{Chauvie}}, \citenamefont {{Chytracek}}, \citenamefont {{Cirrone}},
  \citenamefont {{Cooperman}}, \citenamefont {{Cosmo}}, \citenamefont
  {{Cuttone}}, \citenamefont {{Daquino}}, \citenamefont {{Donszelmann}},
  \citenamefont {{Dressel}}, \citenamefont {{Folger}}, \citenamefont
  {{Foppiano}}, \citenamefont {{Generowicz}}, \citenamefont {{Grichine}},
  \citenamefont {{Guatelli}}, \citenamefont {{Gumplinger}}, \citenamefont
  {{Heikkinen}}, \citenamefont {{Hrivnacova}}, \citenamefont {{Howard}},
  \citenamefont {{Incerti}}, \citenamefont {{Ivanchenko}}, \citenamefont
  {{Johnson}}, \citenamefont {{Jones}}, \citenamefont {{Koi}}, \citenamefont
  {{Kokoulin}}, \citenamefont {{Kossov}}, \citenamefont {{Kurashige}},
  \citenamefont {{Lara}}, \citenamefont {{Larsson}}, \citenamefont {{Lei}},
  \citenamefont {{Link}}, \citenamefont {{Longo}}, \citenamefont {{Maire}},
  \citenamefont {{Mantero}}, \citenamefont {{Mascialino}}, \citenamefont
  {{McLaren}}, \citenamefont {{Mendez Lorenzo}}, \citenamefont {{Minamimoto}},
  \citenamefont {{Murakami}}, \citenamefont {{Nieminen}}, \citenamefont
  {{Pandola}}, \citenamefont {{Parlati}}, \citenamefont {{Peralta}},
  \citenamefont {{Perl}}, \citenamefont {{Pfeiffer}}, \citenamefont {{Pia}},
  \citenamefont {{Ribon}}, \citenamefont {{Rodrigues}}, \citenamefont
  {{Russo}}, \citenamefont {{Sadilov}}, \citenamefont {{Santin}}, \citenamefont
  {{Sasaki}}, \citenamefont {{Smith}}, \citenamefont {{Starkov}}, \citenamefont
  {{Tanaka}}, \citenamefont {{Tcherniaev}}, \citenamefont {{Tome}},
  \citenamefont {{Trindade}}, \citenamefont {{Truscott}}, \citenamefont
  {{Urban}}, \citenamefont {{Verderi}}, \citenamefont {{Walkden}},
  \citenamefont {{Wellisch}}, \citenamefont {{Williams}}, \citenamefont
  {{Wright}},\ and\ \citenamefont {{Yoshida}}}]{1610988}%
  \BibitemOpen
  \bibfield  {author} {\bibinfo {author} {\bibfnamefont {J.}~\bibnamefont
  {{Allison}}} \emph {et~al.},\ }\bibfield  {title} {\bibinfo {title} {Geant4
  developments and applications},\ }\href@noop {} {\bibfield  {journal}
  {\bibinfo  {journal} {IEEE Transactions on Nuclear Science}\ }\textbf
  {\bibinfo {volume} {53}},\ \bibinfo {pages} {270} (\bibinfo {year}
  {2006})}\BibitemShut {NoStop}%
\bibitem [{\citenamefont {Allison}\ \emph {et~al.}(2016)\citenamefont
  {Allison}, \citenamefont {Amako}, \citenamefont {Apostolakis}, \citenamefont
  {Arce}, \citenamefont {Asai}, \citenamefont {Aso}, \citenamefont {Bagli},
  \citenamefont {Bagulya}, \citenamefont {Banerjee}, \citenamefont {Barrand},
  \citenamefont {Beck}, \citenamefont {Bogdanov}, \citenamefont {Brandt},
  \citenamefont {Brown}, \citenamefont {Burkhardt}, \citenamefont {Canal},
  \citenamefont {Cano-Ott}, \citenamefont {Chauvie}, \citenamefont {Cho},
  \citenamefont {Cirrone}, \citenamefont {Cooperman}, \citenamefont
  {Cortés-Giraldo}, \citenamefont {Cosmo}, \citenamefont {Cuttone},
  \citenamefont {Depaola}, \citenamefont {Desorgher}, \citenamefont {Dong},
  \citenamefont {Dotti}, \citenamefont {Elvira}, \citenamefont {Folger},
  \citenamefont {Francis}, \citenamefont {Galoyan}, \citenamefont {Garnier},
  \citenamefont {Gayer}, \citenamefont {Genser}, \citenamefont {Grichine},
  \citenamefont {Guatelli}, \citenamefont {Guèye}, \citenamefont {Gumplinger},
  \citenamefont {Howard}, \citenamefont {Hřivnáčová}, \citenamefont
  {Hwang}, \citenamefont {Incerti}, \citenamefont {Ivanchenko}, \citenamefont
  {Ivanchenko}, \citenamefont {Jones}, \citenamefont {Jun}, \citenamefont
  {Kaitaniemi}, \citenamefont {Karakatsanis}, \citenamefont {Karamitros},
  \citenamefont {Kelsey}, \citenamefont {Kimura}, \citenamefont {Koi},
  \citenamefont {Kurashige}, \citenamefont {Lechner}, \citenamefont {Lee},
  \citenamefont {Longo}, \citenamefont {Maire}, \citenamefont {Mancusi},
  \citenamefont {Mantero}, \citenamefont {Mendoza}, \citenamefont {Morgan},
  \citenamefont {Murakami}, \citenamefont {Nikitina}, \citenamefont {Pandola},
  \citenamefont {Paprocki}, \citenamefont {Perl}, \citenamefont {Petrović},
  \citenamefont {Pia}, \citenamefont {Pokorski}, \citenamefont {Quesada},
  \citenamefont {Raine}, \citenamefont {Reis}, \citenamefont {Ribon},
  \citenamefont {Fira]}, \citenamefont {Romano}, \citenamefont {Russo},
  \citenamefont {Santin}, \citenamefont {Sasaki}, \citenamefont {Sawkey},
  \citenamefont {Shin}, \citenamefont {Strakovsky}, \citenamefont {Taborda},
  \citenamefont {Tanaka}, \citenamefont {Tomé}, \citenamefont {Toshito},
  \citenamefont {Tran}, \citenamefont {Truscott}, \citenamefont {Urban},
  \citenamefont {Uzhinsky}, \citenamefont {Verbeke}, \citenamefont {Verderi},
  \citenamefont {Wendt}, \citenamefont {Wenzel}, \citenamefont {Wright},
  \citenamefont {Wright}, \citenamefont {Yamashita}, \citenamefont {Yarba},\
  and\ \citenamefont {Yoshida}}]{ALLISON2016186}%
  \BibitemOpen
  \bibfield  {author} {\bibinfo {author} {\bibfnamefont {J.}~\bibnamefont
  {Allison}} \emph {et~al.},\ }\bibfield  {title} {\bibinfo {title} {Recent
  developments in geant4},\ }\href
  {https://doi.org/https://doi.org/10.1016/j.nima.2016.06.125} {\bibfield
  {journal} {\bibinfo  {journal} {Nucl. Inst. \& Meth. in Phys. Res. A}\
  }\textbf {\bibinfo {volume} {835}},\ \bibinfo {pages} {186 } (\bibinfo {year}
  {2016})}\BibitemShut {NoStop}%
\bibitem [{\citenamefont {Cleland}\ \emph {et~al.}(1972)\citenamefont
  {Cleland}, \citenamefont {Bailey}, \citenamefont {Eckhause}, \citenamefont
  {Hughes}, \citenamefont {Prepost}, \citenamefont {Rothberg},\ and\
  \citenamefont {Mobley}}]{PhysRevA.5.2338}%
  \BibitemOpen
  \bibfield  {author} {\bibinfo {author} {\bibfnamefont {W.~E.}\ \bibnamefont
  {Cleland}}, \bibinfo {author} {\bibfnamefont {J.~M.}\ \bibnamefont {Bailey}},
  \bibinfo {author} {\bibfnamefont {M.}~\bibnamefont {Eckhause}}, \bibinfo
  {author} {\bibfnamefont {V.~W.}\ \bibnamefont {Hughes}}, \bibinfo {author}
  {\bibfnamefont {R.}~\bibnamefont {Prepost}}, \bibinfo {author} {\bibfnamefont
  {J.~E.}\ \bibnamefont {Rothberg}},\ and\ \bibinfo {author} {\bibfnamefont
  {R.~M.}\ \bibnamefont {Mobley}},\ }\bibfield  {title} {\bibinfo {title}
  {Muonium. {III}. precision measurement of the muonium hyperfine-structure
  interval at strong magnetic field},\ }\href
  {https://doi.org/10.1103/PhysRevA.5.2338} {\bibfield  {journal} {\bibinfo
  {journal} {Phys. Rev. A}\ }\textbf {\bibinfo {volume} {5}},\ \bibinfo {pages}
  {2338} (\bibinfo {year} {1972})}\BibitemShut {NoStop}%
\bibitem [{\citenamefont {Ehrlich}\ \emph {et~al.}(1969)\citenamefont
  {Ehrlich}, \citenamefont {Hofer}, \citenamefont {Magnon}, \citenamefont
  {Stowell}, \citenamefont {Swanson},\ and\ \citenamefont
  {Telegdi}}]{PhysRevLett.23.513}%
  \BibitemOpen
  \bibfield  {author} {\bibinfo {author} {\bibfnamefont {R.~D.}\ \bibnamefont
  {Ehrlich}}, \bibinfo {author} {\bibfnamefont {H.}~\bibnamefont {Hofer}},
  \bibinfo {author} {\bibfnamefont {A.}~\bibnamefont {Magnon}}, \bibinfo
  {author} {\bibfnamefont {D.}~\bibnamefont {Stowell}}, \bibinfo {author}
  {\bibfnamefont {R.~A.}\ \bibnamefont {Swanson}},\ and\ \bibinfo {author}
  {\bibfnamefont {V.~L.}\ \bibnamefont {Telegdi}},\ }\bibfield  {title}
  {\bibinfo {title} {Determination of the muonium hyperfine splitting at low
  pressure from a field-independent zeeman transition},\ }\href
  {https://doi.org/10.1103/PhysRevLett.23.513} {\bibfield  {journal} {\bibinfo
  {journal} {Phys. Rev. Lett.}\ }\textbf {\bibinfo {volume} {23}},\ \bibinfo
  {pages} {513} (\bibinfo {year} {1969})}\BibitemShut {NoStop}%
\bibitem [{\citenamefont {Crane}\ \emph {et~al.}(1971)\citenamefont {Crane},
  \citenamefont {Casperson}, \citenamefont {Crane}, \citenamefont {Egan},
  \citenamefont {Hughes}, \citenamefont {Stambaugh}, \citenamefont {Thompson},\
  and\ \citenamefont {Putlitz}}]{PhysRevLett.27.474}%
  \BibitemOpen
  \bibfield  {author} {\bibinfo {author} {\bibfnamefont {T.}~\bibnamefont
  {Crane}}, \bibinfo {author} {\bibfnamefont {D.}~\bibnamefont {Casperson}},
  \bibinfo {author} {\bibfnamefont {P.}~\bibnamefont {Crane}}, \bibinfo
  {author} {\bibfnamefont {P.}~\bibnamefont {Egan}}, \bibinfo {author}
  {\bibfnamefont {V.~W.}\ \bibnamefont {Hughes}}, \bibinfo {author}
  {\bibfnamefont {R.}~\bibnamefont {Stambaugh}}, \bibinfo {author}
  {\bibfnamefont {P.~A.}\ \bibnamefont {Thompson}},\ and\ \bibinfo {author}
  {\bibfnamefont {G.~z.}\ \bibnamefont {Putlitz}},\ }\bibfield  {title}
  {\bibinfo {title} {{Observation of a Quadratic Term in the hfs Pressure Shift
  for Muonium and a New Precise Value for Muonium}
  $\ensuremath{\Delta}\ensuremath{\nu}$},\ }\href
  {https://doi.org/10.1103/PhysRevLett.27.474} {\bibfield  {journal} {\bibinfo
  {journal} {Phys. Rev. Lett.}\ }\textbf {\bibinfo {volume} {27}},\ \bibinfo
  {pages} {474} (\bibinfo {year} {1971})}\BibitemShut {NoStop}%
\bibitem [{\citenamefont {Shimomura}(2011)}]{doi:10.1063/1.3644324}%
  \BibitemOpen
  \bibfield  {author} {\bibinfo {author} {\bibfnamefont {K.}~\bibnamefont
  {Shimomura}},\ }\bibfield  {title} {\bibinfo {title} {Possibility of precise
  measurements of muonium {HFS} at {J-PARC MUSE}},\ }\href
  {https://doi.org/10.1063/1.3644324} {\bibfield  {journal} {\bibinfo
  {journal} {AIP Conf. Proc.}\ }\textbf {\bibinfo {volume} {1382}},\ \bibinfo
  {pages} {245} (\bibinfo {year} {2011})}\BibitemShut {NoStop}%
\bibitem [{\citenamefont {Kanda}\ \emph {et~al.}(2020)\citenamefont {Kanda},
  \citenamefont {Fukao}, \citenamefont {Ikedo}, \citenamefont {Ishida},
  \citenamefont {Iwasaki}, \citenamefont {Kawall}, \citenamefont {Kawamura},
  \citenamefont {Kojima}, \citenamefont {Kurosawa}, \citenamefont {Matsuda},
  \citenamefont {Mibe}, \citenamefont {Miyake}, \citenamefont {Nishimura},
  \citenamefont {Saito}, \citenamefont {Sato}, \citenamefont {Seo},
  \citenamefont {Shimomura}, \citenamefont {Strasser}, \citenamefont {Tanaka},
  \citenamefont {Tanaka}, \citenamefont {Torii}, \citenamefont {Toyoda},\ and\
  \citenamefont {Ueno}}]{k2020new}%
  \BibitemOpen
  \bibfield  {author} {\bibinfo {author} {\bibfnamefont {S.}~\bibnamefont
  {Kanda}} \emph {et~al.},\ }\href@noop {} {\bibinfo {title} {New precise
  spectroscopy of the hyperfine structure in muonium with a high-intensity
  pulsed muon beam}} (\bibinfo {year} {2020}),\ \Eprint
  {https://arxiv.org/abs/2004.05862} {arXiv:2004.05862 [hep-ex]} \BibitemShut
  {NoStop}%
\bibitem [{\citenamefont {Aoyagi}\ \emph {et~al.}(2020)\citenamefont {Aoyagi},
  \citenamefont {Honda}, \citenamefont {Ikeda}, \citenamefont {Ikeno},
  \citenamefont {Kawagoe}, \citenamefont {Kohriki}, \citenamefont {Kume},
  \citenamefont {Mibe}, \citenamefont {Namba}, \citenamefont {Nishimura},
  \citenamefont {Saito}, \citenamefont {Sasaki}, \citenamefont {Sato},
  \citenamefont {Sato}, \citenamefont {Sendai}, \citenamefont {Shimomura},
  \citenamefont {Shirabe}, \citenamefont {Shoji}, \citenamefont {Suda},
  \citenamefont {Suehara}, \citenamefont {Takatomi}, \citenamefont {Tanaka},
  \citenamefont {Tojo}, \citenamefont {Tsukada}, \citenamefont {Uchida},
  \citenamefont {Ushizawa}, \citenamefont {Wauke}, \citenamefont {Yamanaka},\
  and\ \citenamefont {Yoshioka}}]{Aoyagi_2020}%
  \BibitemOpen
  \bibfield  {author} {\bibinfo {author} {\bibfnamefont {T.}~\bibnamefont
  {Aoyagi}} \emph {et~al.},\ }\bibfield  {title} {\bibinfo {title} {Performance
  evaluation of a silicon strip detector for positrons/electrons from a pulsed
  a muon beam},\ }\href {https://doi.org/10.1088/1748-0221/15/04/p04027}
  {\bibfield  {journal} {\bibinfo  {journal} {J. Instrum.}\ }\textbf {\bibinfo
  {volume} {15}}\bibinfo  {number} { (04)},\ \bibinfo {pages}
  {P04027}}\BibitemShut {NoStop}%
\end{thebibliography}%

\end{document}